\documentclass[secnumarabic, pre, showkeys, showpacs]{revtex4}

\usepackage{bm}
\usepackage{amsmath,amsfonts}
\usepackage{graphicx}
\usepackage[usenames]{color}

\begin{document}

\title{Michelson interferometer operating at effects of first order with respect to v/c\\(\small{the third method of measuring the speed of "aether wind"})}

\author{V.V.Demjanov}
\affiliation{Ushakov State Maritime Academy, Novorossiysk, Russia}
\email{demjanov@nsma.ru}
\date{19 April 2010}

\begin{abstract}

In the first version of this paper (arXiv: 1003.2899v1, 15.03.2010) there is described first, traditional method of measuring the non-zero shift of interference fringe in the Michelson interferometer, operating on the effects of second order with respect to $\upsilon/c$, and are revealed hidden causes of the failure to measure the shift of interference fringe in the period from 1881 till 1960. It is shown that at the latitude of Obninsk within a 24-hour observation period the horizontal projection of aether wind velocity varies from 140 km/s to 480 km/s. The second version of this paper (arXiv: 1003.2899v2, 15.04.2010) is supplemented with a second method of finding the velocity of the aether wind $-$  through measuring the largest seasonal decrease in the ratio of the summer shift of the interference fringe to the winter one (equaled $\sim 12\%$). It gave the same interval of values of the projections of the aether wind velocity as the first method. Below the third method of measuring the aether wind is described that appears to be in agreement with the first two methods.

More than hundred years there persists a belief that Michelson-type interferometer can not be adjusted such as to detect effects of the first order with respect to $\upsilon/c$. Below I show that it is possible to measure the interference fringe shift (and thus the "aether wind") on the first order Michelson interferometer, and more successfully than on the interferometer of the second order. In contrast to the traditional approach, in the interferometer of the first order the light after splitting on a semi-transparent plate propagates in both arms to the reflecting mirrors in one optical medium (with dielectric permittivity $\varepsilon_1$), and returns after reflection from the mirrors through  another optical medium (with  dielectric permittivity $\varepsilon_2$). The shift of the interference fringe is reliably detected in the experiment when turning the interferometer by $90^\circ$. It was found to be proportional to $\varepsilon_1-\varepsilon_2$.

 Experimental data are interpreted in the bounds of the Fresnel drag of light by a moving optical medium neglecting terms quadratic in $\mathit{\upsilon}/c$. The horizontal projection $\mathit{\upsilon}$  of the Earth's velocity relative to luminiferous aether thus found lies in the range 140 km/s $<\mathit{\upsilon}<$ 480 km/s depending on the time of the day and night at the latitude of Obninsk. This is the third method of measuring the speed of aether wind. It gives the same range of values as two earlier described methods operating at second order with respect to $\mathit{\upsilon}/c$.
\end{abstract}
\maketitle

\section{Two variants of aether wind's detector}

As is known \cite{Demjanov,Demjanov rus} there is no fringe shift in evacuated Michelson interferometer. Thus, this mode of the device is not fit to detect ''aether wind''. Michelson interferometer becomes sensitive to ''aether wind'' when using in it as light's carrier an optical medium with refractive index $n>1$ \cite{Demjanov rus,Demjanov}. If light passes always in a single medium the interference fringe shift $X_m$ appears to be proportional to square of the velocity $\mathit{\upsilon}$ of the interferometer relative to aether: $X_m\sim(\mathit{\upsilon}/c)^2$, where $\mathit{\upsilon}\ll c$. Here the allowance for Lorentz contraction of the interferometer's longitudinal arm is vital.

Below I report results of the experiment on Michelson interferometer configured in such a way that light runs consecutively via two different optical media $-$ it goes at first through one medium and then returns back through other medium. In this case the fringe shift proves to be proportional to $\mathit{\upsilon}/c$ where
$\mathit{\upsilon}\ll c$. So, the two-media-mode device is more sensible to ''aether wind''. Besides, in the interferometer of the first order the contribution of relativistic  factor into the fringe shift, being quadratic with respect to $\mathit{\upsilon}/c$, can be neglected as  small quantity.

\section{Two-media detector of aether wind}

\begin{figure}[h]
  \begin{center}
 \includegraphics[scale=0.6]{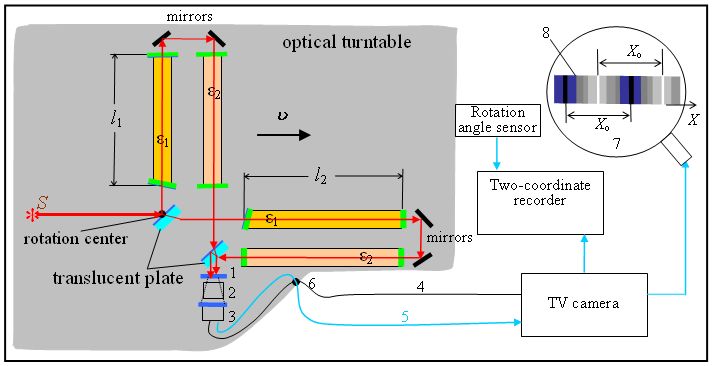}
  \caption{Functional scheme of the first order by $\mathit{\upsilon}/c$ interferometer (1971 year). Light issuing from the source $S$ splits at the semi-transparent plate and then goes consecutively through two glass tubes filled with light's carriers of respective dielectric permittivity $\varepsilon_1$ and $\varepsilon_2$. Ends of the tubes are covered by thin glass lids. After reflecting from  rebounding mirrors the light returns for interference at the other half-transparent plate. 1 is the interference screen, 2 telescopic ocular, 3 vidicon with a deflecting facility, 4 and 5 are feeding and video cords passing via the tube 6 in the center of rotation, 8 is the interference pattern at the screen of the kinescope 7.}\label{fig1}
\end{center}
\end{figure}

The device in question is a common rotary cross-like interferometer (Fig.\ref{fig1}). Light firstly splits by the half-transparent plate into two orthogonal beams. Further each beam goes to rebounding mirrors through the optical medium having dielectric permittivity $\varepsilon_1$ and then returns in the medium having dielectric permittivity  $\varepsilon_2$. Media are separated in space by minor displacement of return containers. The respective shear of light's beams is secured by means of two mirrors. Thus,  returning rays meet in the point laying at another semi-transparent plate that is angled at $45^\circ$ just as the splitting plate. The space separation  from the incoming beam of the light outgoing to the interference screen (1 on Fig.\ref{fig1})  brings down parasitic noise as compared with usual construction of the second order interferometer. Next, collected on the plate 1 (Fig.1) rays are projected by the telescopic objective 2 on the screen of the vidicon 3 and then transmitted by the TV camera on the screen of the kinescope 7.

\section{Computing aether wind velocity from interference fringe shift}

The interference fringe shift is proportional to difference $\Delta t$ of round-trip times according to
\begin{equation}
X_m=c\frac{X_o}{\lambda}\Delta t\label{fringe shift}
\end{equation}
where $X_o$ is the bandwidth and $\lambda$ the wavelength. By virtue of (\ref{fringe shift}) we may reason in terms of time intervals $t$.

The speed of light $\tilde{c}$ in the moving with velocity $\mathit{\upsilon}>0$ optical medium is given by Fresnel formula
\begin{equation}
\tilde{c}'_\pm=\frac{c}{n}\pm \mathit{\upsilon}(1-\frac{1}{n^2})=\frac{c}{\sqrt\varepsilon}\pm \mathit{\upsilon}\frac{\Delta\varepsilon}{\varepsilon}\label{Fresnel}
\end{equation}
as measured in the reference frame of the moving medium. Here $n$ is the refractive index and $\Delta\varepsilon=n^2-1$ the contribution of particles into the dielectric permittivity $\varepsilon=n^2$ of the optical medium.

From (\ref{Fresnel}) we have for the round-trip time $t_\parallel$ in the direction of interferometer arm parallel to $\mathit{\upsilon}$
\begin{equation}
t_\parallel=\frac{l'}{\tilde{c}'_+}+\frac{l'}{\tilde{c}'_-}=\frac{l'}{c/{\sqrt\varepsilon_1}+\mathit{\upsilon}\Delta\varepsilon_1/\varepsilon_1}+\frac{l'}{c/{\sqrt\varepsilon_2}-\mathit{\upsilon}\Delta\varepsilon_2/\varepsilon_2}
\approx\frac{l}{c}\left[{\sqrt\varepsilon_1}+{\sqrt\varepsilon_2}+\frac{\mathit{\upsilon}}{c}(\Delta\varepsilon_1-\Delta\varepsilon_2)\right]\label{time parallel}
\end{equation}
where $\varepsilon_1$ is the dielectric permittivity of the forward medium, and $\varepsilon_2$ that of the return medium. In deriving (\ref{time parallel}) terms with $(\mathit{\upsilon}/c)^2$ were discarded.

We have for the round-trip time in the orthogonal direction
\begin{equation}
t_\perp=\frac{l}{\sqrt{(c/n_1)^2+\mathit{\upsilon}^2}}+\frac{l}{\sqrt{(c/n_2)^2+\mathit{\upsilon}^2}}\approx\frac{l}{c}({\sqrt\varepsilon_1}+{\sqrt\varepsilon_2})\label{time perpendicular}
\end{equation}
where again terms with $(\mathit{\upsilon}/c)^2$ were neglected.

Subtracting (\ref{time parallel}) from (\ref{time perpendicular}) gives
\begin{equation}
\Delta t=t_\perp-t_\parallel\approx\frac{\upsilon}{c}\frac{l}{c}(\Delta\varepsilon_1-\Delta\varepsilon_2)=\frac{\upsilon}{c}\frac{l}{c}(\varepsilon_1-\varepsilon_2).\label{time difference}
\end{equation}
Formula (\ref{time difference}) shows that the difference of round-trip times depends on the difference of the contribution of particles into dielectric permittivities of two light carrying media.

\section{Measuring interference fringe shift}

\begin{figure}[h]
  \begin{center}
 \includegraphics[scale=0.5]{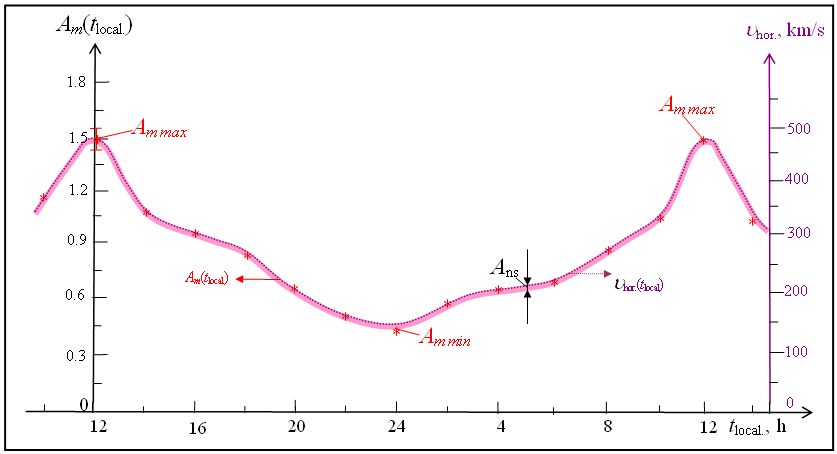}
  \caption{Measured relative amplitude $A_m=X_m/X_o$ at various times of day and night, local time at Obninsk 55.8$^\circ$ NL, 22 June 1971 year, where $X_m$ is the amplitude of interference fringe shift and $X_o=90$ mm is the bandwidth on the kinescope's screen. Light's carriers are CS$_2$ having $\varepsilon\approx0.0036$ and air having $\varepsilon\approx0.0006$. The line width shows the  jitter noise $A_{ns}$ of the interferometer. The horizontal projection $\mathit{\upsilon}_{hor.}$ of the aether wind velocity, computed from $A_m$ by (\ref{time difference}) with  (\ref{fringe shift}) is shown at the right ordinate axis. Parameters of the experiment: arms are $l_1=l_2=0.2$ m, wavelength $\lambda=6\cdot10^{-7}$ m.}\label{fig2}
\end{center}
\end{figure}

Fig.\ref{fig2} shows the measured amplitude $A_m$ of the harmonic component of the interference fringe shift as a function of local time $t_{local}$ at the attitude of Obninsk. Measurements covered 24 hours of the day and night. The light's carriers were: in the forward direction carbon bisulfide (CS$_2$) $\Delta\varepsilon_1=0.0036$, and on the return path the air at normal pressure,  $\Delta\varepsilon_1=0.0006$. The right scale of ordinates shows values of the aether wind velocity $\mathit{\upsilon}_{hor.}\sim{A_m}$ km/s. The dependence $\mathit{\upsilon}(t_{local})$ agrees well with that obtained on the  interferometer of second order with respect to $\upsilon/c$ \cite{Demjanov rus,Demjanov}.

According to (\ref{time difference}) the first order device is $(\upsilon/c)^{-1}\approx1000$ times more sensitive than  the interferometer of the second order with respect to $\upsilon/c$. This enabled me to use a relatively small arms $l=0.2$ m.

\begin{figure}[h]
  \begin{center}
\includegraphics[scale=0.5]{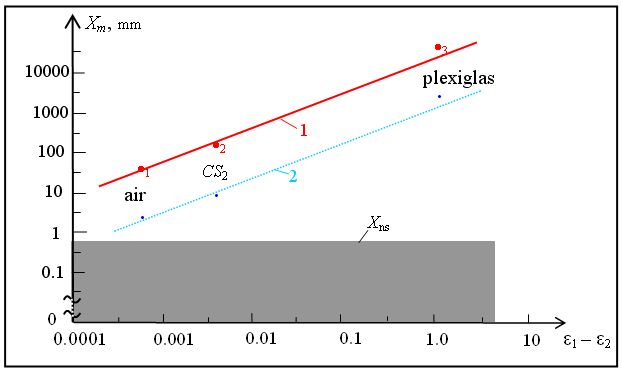}
  \caption{Measured shift $X_m$ of the interference fringe as a function of difference $\varepsilon_1-\varepsilon_2$ between dielectric permittivities of two light's carriers: ${\color{red}\bullet}_1$  air ($\varepsilon_1=1.0006$)/lab vacuum ($\varepsilon_2=1.000006$), ${\color{red}\bullet}_2$  CS$_2$ ($\varepsilon_1=1.0036$)/air, ${\color{red}\bullet}_3$ plexiglas ($\varepsilon_1\approx2$)/air. Abscissa and ordinate scales are logarithmic. Line 1 corresponds to $X_{m\,\,max}$, line 2 to $X_{m\,\,min}$ in notations of Fig.\ref{fig2}. $\delta X_{ns}$ is the jitter noise read on the screen of kinescope. Parameters of the experiment for CS$_2$: arms are $l_1=l_2=0.2$ m, wavelength $\lambda=6\cdot10^{-7}m$ (all other data are reduced to this values)}\label{fig3}
\end{center}
\end{figure}

Fig.\ref{fig3} shows the dependence of the interference fringe shift $X_m$ on the difference $\varepsilon_1-\varepsilon_2$ between dielectric permittivities of light's carriers. As we see, $X_m$ linearly grows from $\varepsilon_1-\varepsilon_2=0$ to $\varepsilon_1-\varepsilon_2=1$.  By (\ref{time difference}) the sensitivity can be enhanced in $10^5$ times (100 times due to $\varepsilon_1-\varepsilon_2$ and 1000 times due to $\upsilon/c$). So that the measuring can be performed at microwaves,  i.e. $10^5$ times greater wavelengths than in optics. The microwave  experiment was made using the CaTiO$_3$ ferroelectric guide, $\varepsilon = 255$, at $\lambda=10$ cm by the direct measurement of the phase difference. The converted to $X_m$ magnitude lies exactly on the continuation of the line 1 in Fig.\ref{fig3} \cite{Demjanov rus}.

\section{Conclusion}

Thus the reality of the first order with respect to $\upsilon/c$ Michelson interferometer is experimentally demonstrated. Experiments show that the shift of the interference fringe occurs in optical media and is absent in vacuum. The first order interferometer is $(\upsilon/c)^{-1}\approx1000$ times more sensitive to aether wind than the second order Michelson interferometer. The first order interferometer shows the ratio signal/noise $\sim100$ all round the clock. Such the resolving power and stability of measurements at any time of day and night and any season attests confidently the presence of the absolute reference frame with the Earth moving relative to it with the velocity that is above 480 km/s.

Earlier the same estimation of the ''aether wind'' velocity has been obtained on Michelson interferometer by two methods \cite{Demjanov exp} operating at second order effects with respect to ${\mathit\upsilon}/c$ \cite{Demjanov}. It was the historically first method that took into account parameters of optical medium specially used as a light's carrier. The content presented here is actually a third method of experimental measuring the velocity of ''aether wind'' via the shift of interference fringe at the transverse beams interferometer $-$ the method operating at first order effects with respect to ${\mathit\upsilon}/c$ due to use of two optical media as light's carriers.

Thus, kinetic evidences for existence of luminiferous aether are to this time provided by three phase interferometry techniques, two operating at second order effects and third one, described in the present report, at the first order one.

\begin{acknowledgments}
The author is grateful to Dr V.P.Dmitriyev for valuable comments, fruitful discussion and assistance in the preparation of this manuscript.
\end{acknowledgments}

\end{document}